\newcommand{\BaCdV} {BaCdVO(PO$_4$)$_2$\xspace}

\documentclass[aps,pra,reprint,showpacs,longbibliography,superscriptaddress, citeonline]{revtex4-1}
\usepackage{amsmath,amssymb,bm}
\usepackage{graphicx}
\usepackage{xspace}
\usepackage{latexsym}
\usepackage{color}
\usepackage{hyperref}
\usepackage{placeins}

\begin{document}
\title{Thermodynamics of
a frustrated quantum magnet on a square lattice}

\author{K.~Yu.~Povarov}
    \email{povarovk@phys.ethz.ch}
    \affiliation{Laboratory for Solid State Physics, ETH Z\"{u}rich, 8093 Z\"{u}rich, Switzerland}
    \homepage{http://www.neutron.ethz.ch/}

\author{V.~K.~Bhartiya}
    \affiliation{Laboratory for Solid State Physics, ETH Z\"{u}rich, 8093 Z\"{u}rich, Switzerland}

\author{Z.~Yan}
    \affiliation{Laboratory for Solid State Physics, ETH Z\"{u}rich, 8093 Z\"{u}rich, Switzerland}

\author{A.~Zheludev}
    \affiliation{Laboratory for Solid State Physics, ETH Z\"{u}rich, 8093 Z\"{u}rich, Switzerland}

\begin{abstract}
We report the magnetic and calorimetric measurements in {\em single
crystal} samples of the square lattice $J_{1}-J_{2}$  quantum
antiferromagnet \BaCdV. An investigation of the scaling of
magnetization reveals a ``dimensionality reduction'' indicative of a
strong degree of geometric frustration. Below a characteristic
temperature of $T^{\ast}\simeq150$~mK we observe the emergence of an
additional strongly fluctuating quantum phase close to full magnetic
saturation. It is separated from the magnetically ordered state by
first- and second-order phase transitions, depending on the
orientation of the applied magnetic field. We suggest that this
phase may indeed be related to the theoretically predicted
spin-nematic state.
\end{abstract}

\date{\today}
\maketitle

\section{Introduction}

\begin{figure}
    \includegraphics[width=0.5\textwidth]{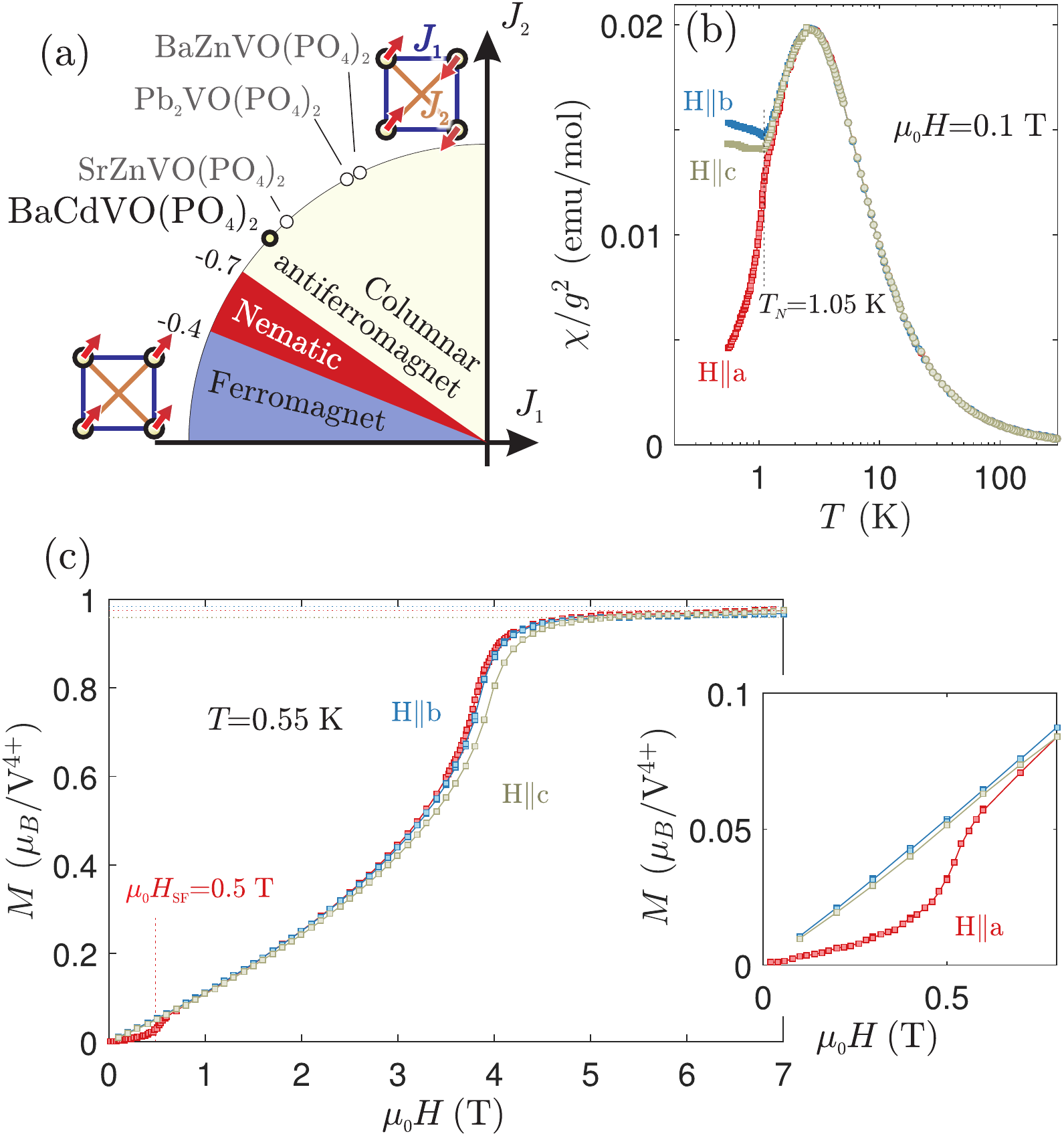}\\
    \caption{(a) The frustrated $S=1/2$ Heisenberg square lattice model and typical ``circular'' representation of its ground
        state as a function of the ferromagnetic $J_{1}$ to antiferromagnetic $J_{2}$ exchange ratio.
         The positions of
        \BaCdV\ as well as of few other similarly structured materials
        are shown (after Ref.~\cite{NathTsirlin_PRB_2008_BaCdVO(PO4)2}).
        (b) Magnetic susceptibilities along the different directions of \BaCdV\ crystal,
        scaled with their $g-$factors.
        (c) Isothermal magnetization along different directions at
        $T=0.55$~K. The inset shows the region around
        $\mu_{0}H_{\text{SF}}\simeq0.5$~T in more detail. The dashed lines show the estimate for $M_{\text{sat}}=g\mu_{B}/2$ value.
        All the magnetic data is corrected for the diamagnetic
        background.
    }\label{FIG:Magnetic}
\end{figure}

The quest for the so-called spin-nematic state in magnetic
insulators started more than three decades ago, but continues to
this
day~\cite{AndreevGrishchuk_JETP_1984_SpinNematics,Chubukov_PRB_1991_Chains,ShannonMomoi_PRL_2006_J1J2squarecircle,ZhitomirskyTsunetsugu_EPL_2010_nematic,
ButtgenNawa_PRB_2014_LiCuVO4yetanothernematic,Orlova_PRL_2017_moreNematicLiCuVO4}.
This exotic magnetic order spontaneously breaks rotational symmetry,
while keeping time-reversal symmetry intact. It can be understood as
a quantum condensate of bound magnon
pairs~\cite{Chubukov_PRB_1991_Chains,ShannonMomoi_PRL_2006_J1J2squarecircle,ZhitomirskyTsunetsugu_EPL_2010_nematic}.
The key characteristics of any potential host system are competing
ferro- (FM) and antiferromagnetic (AF) interactions and extreme
quantum fluctuations. The baseline model is  the $S=1/2$ square
lattice Heisenberg Hamiltonian with FM nearest-neighbor exchange
$J_1$ and AF next-nearest-neighbor  coupling
$J_2$~\cite{Shannon_EPJB_2004_GenericJ1J2theory,ShannonMomoi_PRL_2006_J1J2squarecircle,ShindouMomoi_PRB_2009_J1J2nematicSlavebosons,Shindou_PRB_2011_J1J2nematicProjective,SchmidtThalmeier_PhysRep_2017_2Dfrustratedreview}
sketched in Fig.~\ref{FIG:Magnetic}(a) alongside its phase diagram.
The classical critical point at $J_{2}/J_{1}=-1/2$ separates
 FM and columnar-AF states, but becomes
destabilized by quantum fluctuations and is replaced by a novel
region in its vicinity. The resulting state can be understood as a
magnon bound pair condensate occurring at zero field
--- the spin-nematic~\cite{ShannonMomoi_PRL_2006_J1J2squarecircle,Shindou_PRB_2011_J1J2nematicProjective}.
Even outside the narrow $J_{2}/J_{1}$ parameter range in which spin
nematic is supposed to exist at zero field, this state can be
further stabilized in a magnetized system. Magnon pair condensation
and hence spin nematicity can be induced by an external magnetic
field close to the saturation point. This result turns out to hold
well away from optimal parameter set $J_{2}/J_{1}=-1/2$ and even in
the presence of additional couplings in the
Hamiltonian~\cite{ShannonMomoi_PRL_2006_J1J2squarecircle,Ueda_JPSJ_2015_NematicInField}.

Despite the vast body of theoretical work, experimentally the spin
nematic state on a frustrated square lattice remains elusive. One
obvious problem is that the tensorial order parameter is invisible
to most conventional magnetism probes. What is an even bigger
obstacle, is that potential model compounds are few and hard to
synthesize. The most promising known candidate is
\BaCdV~\cite{NathTsirlin_PRB_2008_BaCdVO(PO4)2,Tsirlin_PRB_2009_FSQLmagnetization}.
The applicability of the  $J_{1}-J_{2}$ model to this compound has
been validated by density functional theory calculations
~\cite{TsirlinRosner_PRB_2009_FSQLvanadatesSummary}. The material
features strong geometric frustration ($J_{2}/J_{1}\simeq-0.9$) and
easily accessible energy scales (saturation field about $4$~T, AF
order below $T_{N}=1.05$~K).  The high-temperature thermodynamics is
consistently described by $J_{1}=-3.6$~K and
$J_{2}=3.2$~K~\cite{NathTsirlin_PRB_2008_BaCdVO(PO4)2}. For lack of
other candidates, \BaCdV\ has been a subject of intense theoretical
studies, including specific predictions for inelastic neutron
scattering~\cite{Smerald_PRB_2015_INSnematic} and nuclear magnetic
resonance~\cite{SmeraldShannon_PRB_2016_nematicNMR}.
Disappointingly, a lack of single crystals has severely impeded
experimental studies. To date, no empirical evidence of a
spin-nematic phase or any related unconventional magnetism has been
reported in this material.

In the present paper we describe the unusual magnetic and
thermodynamic properties of \emph{single-crystal} samples of \BaCdV.
We map out the anisotropic magnetic phase diagram and study the
``dimensionality reduction'' and peculiar scaling of magnetization
near the field-induced quantum phase transition. We accomplish this
by employing magnetization, specific heat, and the magnetocaloric
effect studies.  In what may be the first sign of spin nematicity,
we report evidence of an additional low-temperature field-induced
strongly fluctuating quantum regime just below saturation. In an
axially symmetric geometry the new state emerges in a first-order
transition, and is preceded by substantial precursor transverse
fluctuations in the magnetically ordered state.

\section{Experiment details}

\subsection{Material}

High-quality single crystals of \BaCdV\ were grown using the
self-flux Bridgman method from the melt of presynthesized
BaCdP$_2$O$_7$ and vanadium dioxide at $1000^{\circ}$~C. The details
of the method will be published elsewhere. The crystal structure
[orthorhombic $P_{bca}$ ($D^{15}_{2h}$, No $61$), $a=8.84$,
$b=8.92$, $c=19.37$~\AA] was validated using single-crystal x-ray
diffraction on a Bruker APEX-II instrument, and found to be totally
consistent with that reported
previously~\cite{Meyer_ZNatur_1997_CrystVanadates}.

\begin{figure}
  \includegraphics[width=0.3\textwidth]{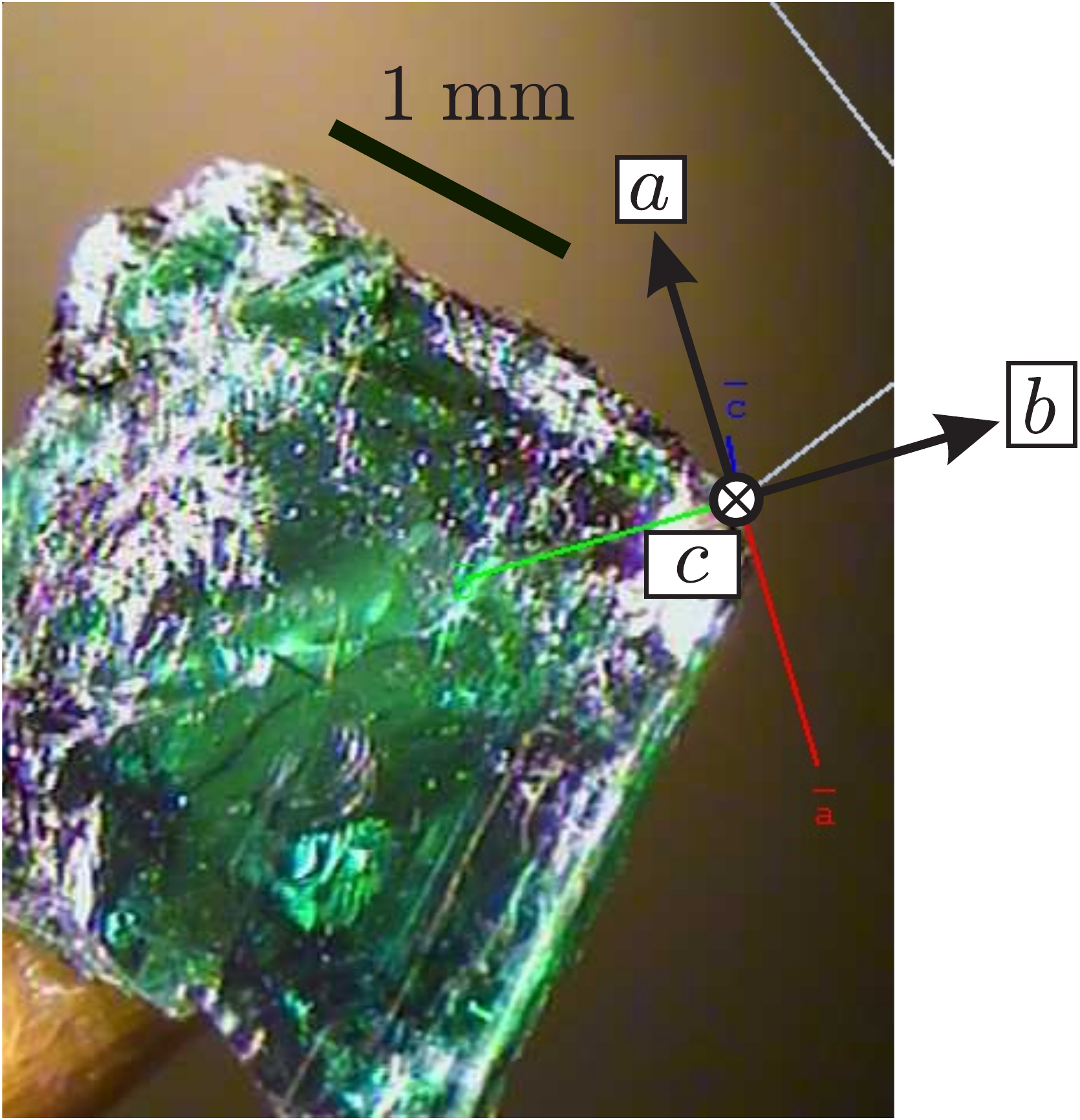}\\
  \caption{A single 16-mg crystal sample of \BaCdV\
 used in magnetization measurements. A snapshot from a Bruker APEX II single-crystal x-ray diffractometer.}\label{FIG:xtalpic}
\end{figure}

The single crystals of \BaCdV\ have the appearance of green
transparent square plates. The plate corresponds to the
crystallographic $ab$ plane, with the directions of $a$ and $b$ axes
being usually along the diagonals. Correspondingly, the $c$ axis is
normal to the plane. One of the crystals from the present study is
shown in Fig.~\ref{FIG:xtalpic}.

We would like to note that the highly symmetric shape of the samples
ensures that no crystal morphology-related effects can be expected
to make a difference between $\mathbf{H}\parallel\mathbf{a}$ and
$\mathbf{H}\parallel\mathbf{b}$ configurations.

\subsection{Techniques}

The magnetization measurements were performed with the $7$~T SQUID
magnetometer [Quantum Design Magnetic Property Measurement System
(MPMS)] in the temperature range $1.8-300$~K. Further extension to
the temperatures of about $0.5$~K was achieved with the help of
$^3$He cryostat inset iQuantum iHelium3. The magnetic susceptibility
$\chi=M/H$ was measured at a small field $0.1$~T. The crystal shown
in Fig.~\ref{FIG:xtalpic} was used in all the magnetic measurements.

Specific heat measurements were carried out on a standard Quantum
Design relaxation calorimetry option and the $^3$He-$^4$He dilution
refrigerator inset for Quantum Design Physical Properties
Measurement System (PPMS). Two measurement geometries
$\mathbf{H}\parallel\mathbf{a}$ and $\mathbf{H}\parallel\mathbf{b}$
were realized by mounting a $2.3$-mg flat single-crystal sample on a
small silver foil holder with Apiezon N grease. The measurement
procedure consists of giving a gentle heat pulse to the sample
platform. Then the temperature rise is observed during the pulse,
and subsequent temperature fall is observed as the heater is turned
off. The resulting $T(t)$ curve typically has a characteristic
``shark fin'' shape, and specific heat can be calculated from the
curvature.

The magnetocaloric effect measurements were performed in the same
setup by directly reading the resistivity of the sample thermometer
as the function of slowly varying magnetic field. It was done either
with an Agilent E4980A LCR meter, or with a Stanford Research SR830
lock-in amplifier.

\section{Results and discussion}
\subsection{Magnetization studies}

\subsubsection{Susceptibility above $T_N$}

Above $T_N$ the susceptibilities [Fig.~\ref{FIG:Magnetic}(b)] show
qualitatively identical behavior: typical Curie--Weiss tail at high
temperatures, followed by a rounded maximum at $T\simeq2.5$~K and
then gradual decrease down to $T_{N}=1.05$~K marked by a kink.

The Curie--Weiss part of the susceptibility curve at high
temperatures can be used for accurate determination of the $g$
factors and the diamagnetic background:

\begin{equation}\label{EQ:CWfit}
    \chi_{\alpha}(T)=\chi_{\alpha}^{0}+\frac{(g_{\alpha}/2)^{2}C}{T+\Theta},
\end{equation}

where $C=0.375$~K~emu/mol is the Curie constant for the $S=1/2$ case
with $g=2.00$. The analysis given by Eq.~(\ref{EQ:CWfit}) was
performed in a temperature window between $30$ and $200$ K. The
Curie--Weiss temperature is found to be $\Theta=-0.90(2)$~K; we have
enforced the equal value for all three directions. The values of $g$
factors and diamagnetic background susceptibilities
$\chi_{\alpha}^{0}$ are summarized in Table~\ref{TAB:CWfit}. The
obtained $g-$factor values are rather isotropic and consistent with
powder EPR estimates~\cite{Foerster_2011_PhDthesis}. In
Fig.~\ref{FIG:Magnetic}(b) one can see that background subtracted
susceptibilities normalized by $g^2$ show a perfect overlap in
absence of magnetic order.

\begin{table}
\begin{tabular}{ c r c r c }
  \hline\hline
    Direction & &$g_\alpha$ & &$\chi^{0}_{\alpha}$ (emu/mol) \\
    \hline
  $\mathbf{a}$ & &1.95(1)& & $-1.79(2)\times10^{-4}$\\
  $\mathbf{b}$ & &1.97(1)& & $-2.45(2)\times10^{-4}$\\
  $\mathbf{c}$ & &1.92(1)& & $-2.32(2)\times10^{-4}$\\
  \hline\hline
\end{tabular}
\caption{Results of the Curie--Weiss analysis of the
high-temperature susceptibility.}\label{TAB:CWfit}
\end{table}

\subsubsection{Easy-axis anisotropy}

Below the N\'{e}el temperature $T_{N}=1.05$~K the susceptibilities
shown in Fig.~\ref{FIG:Magnetic}(b) start to show rather different
behavior. At low temperatures $\chi_{b}(T)$ and $\chi_{c}(T)$ remain
more or less constant, while $\chi_{a}(T)$ shows a rapid decrease
upon cooling. This suggests a collinear magnetic structure with
spins along the $\mathbf{a}$ axis. This interpretation is backed by
isothermal magnetization $M(H)$ scans at $T=0.55$~K
[Fig.~\ref{FIG:Magnetic}(c)]. For the $\mathbf{H}\parallel
\mathbf{a}$ case (and only for that geometry) there is a pronounced
magnetization jump around $\mu_{0}H_{\text{SF}}\simeq0.5$~T. This
behavior is characteristic of a spin-flop transition driven by the
weak Ising-like anisotropy, $\mathbf{a}$ being the magnetic easy
axis. Thus, using a simple Heisenberg model to describe the system
can only be done with caution. Below we shall refer to experiments
with $\mathbf{H}\parallel\mathbf{a}$ as the \emph{axial} geometry,
and to those with a field in perpendicular directions as
\emph{transverse}.

\subsubsection{Convex shape and ``dimensionality reduction''}

\begin{figure}
  \includegraphics[width=0.5\textwidth]{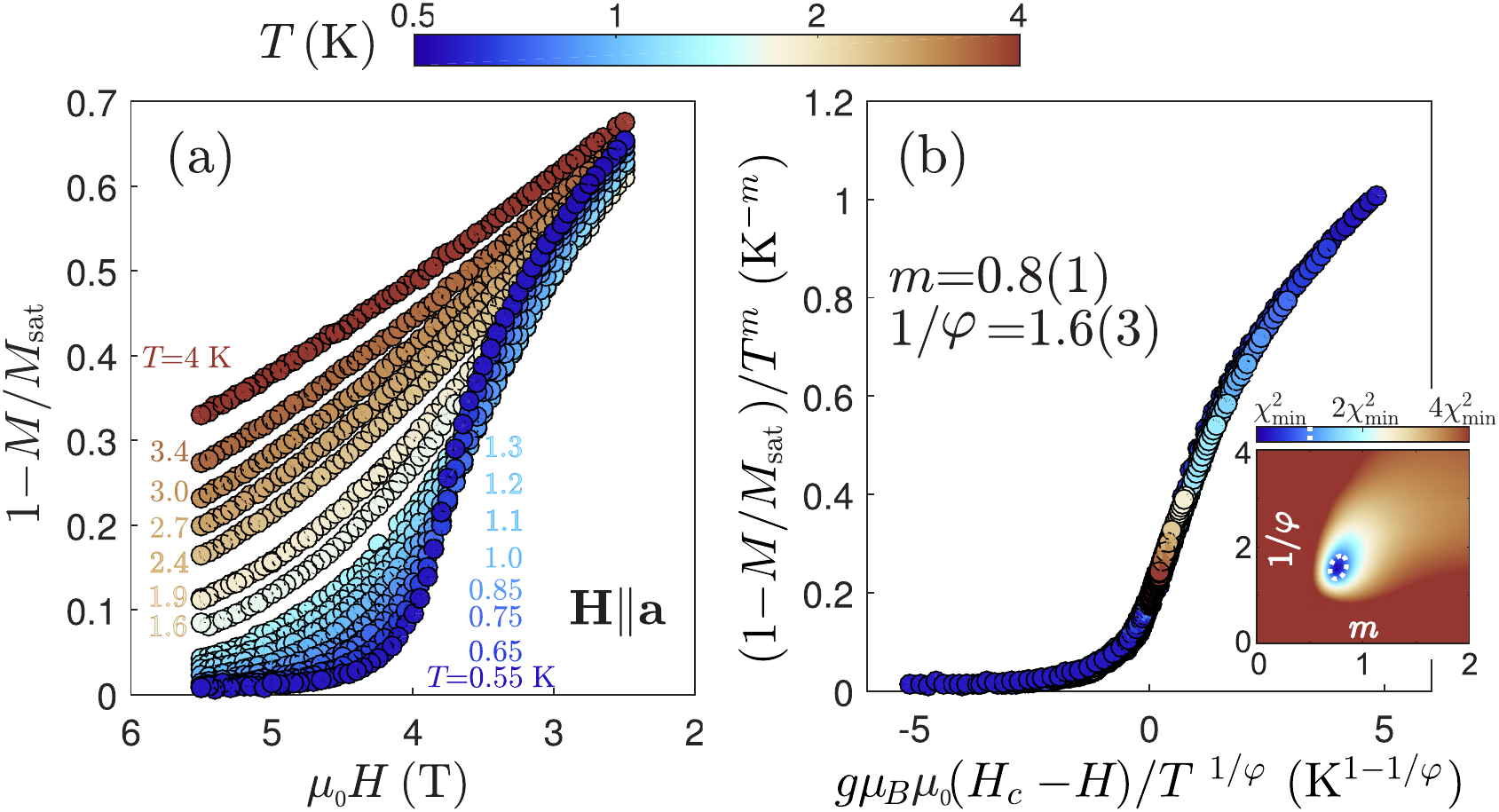}\\
  \caption{Scaling of magnetization, observed near the saturation field ($\mathbf{H}\parallel\mathbf{a}$ case).
  (a) Raw data, taken in the interval $0.5-4$~K.
  (b) Same data, scaled according to Eq.~(\ref{EQ:MHscaling}). The observed exponents are $1/\varphi=1.6(3)$ and
  $m=0.8(1)$.  The inset shows the empirical $\chi^2$ goodness of overlap with highlighted boundary at which the optimal value increases by 50\%.}
  \label{FIG:Mscale}
\end{figure}

The most striking feature of the measured $M(H)$ curves is their
extreme convex shape close to the saturation. As known from the
numerical studies of the $J_{1}-J_{2}$
model~\cite{ThalmeierZhitomirsky_PRB_2008_J1J2magnetization,Tsirlin_PRB_2009_FSQLmagnetization},
it serves as a reliable indicator of the significant magnetic
frustration, indirectly confirming the nearly critical positioning
of \BaCdV\ on the Fig.~\ref{FIG:Magnetic}(b) phase diagram.

We note that the measured convex magnetization curve is reminiscent
of the cusp singularity occurring at saturation in the AF spin
chains~\cite{JeongRonnow_PRB_2015_CupzNscaling,Breunig_SciAdv_2017_CupzNscaling}.
This feature --- square root cusp at the saturation magnetization
$M_{\text{sat}}-M(H)\propto\sqrt{H_{c}-H}$ --- is endemic to one
dimension, yet it appears in our essentially two-dimensional (2D)
material. This is another signature of the frustration, known as the
``dimensionality reduction''. For example, a similar effect is
responsible for low-temperature crossover to effectively 2D behavior
in a nominally 3D material BaCuSi$_2$O$_6$ (``Han
purple'')~\cite{SebastianHarrison_Nature_2006_HanPurpleDimRed}. In
the case of the frustrated square lattice a qualitative prediction
is given by Jackeli and
Zhitomirsky~\cite{JackeliZhitomirsky_PRL_2004_FrustratedBEC}: close
to the points of perfect frustration $|J_{2}/J_{1}|=1/2$ 1D-like
behavior with a square root magnetization cusp is indeed present in
a 2D material at saturation. A simple explanation is, close to $H_c$
the low-energy part of the spin-wave spectrum defining the low-$T$
behavior features a \emph{continuous circle of degenerate minima} as
the result of strong frustration. This effectively reduced the
problem to a one-dimensional one, rendering the low-energy spectrum
as being pseudo-1D~\cite{JackeliZhitomirsky_PRL_2004_FrustratedBEC}.

Thus, verifying the zero temperature
$M_{\text{sat}}-M(H)\propto\sqrt{H_{c}-H}$ prediction would be a
strong signature of nearly critical $J_{2}/J_{1}$ coupling ratio in
\BaCdV. However, in a realistic experiment we are dealing with the
finite temperatures that make the cusp rounded and hide away the
associated power law. Below we will show that by considering the
quantum critical behavior of longitudinal magnetization close to
$H_c$ it is nonetheless possible to relate the ``hidden'' zero
temperature cusp to the available finite temperature data.

\subsubsection{Quantum critical scaling: Theory}

 The basic assumption that we need to make is that the
hyperscaling holds at the quantum critical point. This would be the
case if the ``dimensionality reduction'' scenario takes place
indeed. Then in the vicinity of the transition the free energy can
be expressed as

\begin{equation}
\label{EQ:FscalingSM}
    F(T,H)=\lambda^{b}\mathcal{F}\left[\lambda^{z}T,\lambda^{1/\nu}(H-H_{c})\right]+F_{0}(T,H).
\end{equation}

Here $\mathcal{F}(x,y)$ is some \textit{a priori} unknown function
of two variables and $\lambda$ is an arbitrary positive number. This
term reflects the singular part of the free energy. We do not even
need to make any assumption about the particular value of the
exponent $b$ (which is usually set to be $d+z$ --- the effective
dimensionality of the quantum phase transition). The nonsingular
part of free energy $F_{0}(T,H)$ is important at $H\gg H_{c}$ and
can be approximated as $-M_{\text{sat}}(H-H_{c})$. Then, one can
express the magnetization reduction as:

\begin{align}
\label{EQ:dFdHscalingSM}
    -M_{\text{sat}}+M(T,H)=-\left(\frac{\partial (F-F_{0})}{\partial
    H}\right)_{T}\\\nonumber
    =-\lambda^{b+1/\nu}\mathcal{M}_{0}\left[\lambda^{z}T,\lambda^{1/\nu}(H_{c}-H)\right].
\end{align}

Once again, $\mathcal{M}_{0}(x,y)$ is the unknown function of two
variables. Now, at finite temperatures by setting $\lambda=T^{-1/z}$
one arrives at the following general scaling relation:

\begin{equation}
\label{EQ:MHscaling}
    1-M(H,T)/M_{\text{sat}}=T^{m}\mathcal{M}\left(\frac{g\mu_{B}\mu_{0}(H-H_{c})}{T^{1/\varphi}}\right).
\end{equation}

Here $\varphi=\nu z$ is the \emph{crossover exponent} describing the
interplay between the thermal and quantum fluctuations in the
transition vicinity. The second exponent $m$ also has a simple
physical meaning. It describes the temperature dependence of
magnetization reduction at $H=H_{c}$.

The exponents $\varphi$ and $m$ are also crucial for characterizing
the $T=0$ behavior. To see this, one needs to set
$\lambda=(H_{c}-H)^{-\nu}$ in Eq.~(\ref{EQ:dFdHscalingSM}). Then the
zero-temperature magnetization cusp is described as

\begin{equation}
\label{EQ:MHcusp}
    1-M(H)/M_{\text{sat}}\propto(H_{c}-H)^{m\varphi}.
\end{equation}

So this is the $m\varphi$ product that defines the low-temperature
``cusp singularity'', and this is the quantity that needs to be
found experimentally in order to verify the ``dimensionality
reduction'' prediction by Jackeli and
Zhitomirskii~\cite{JackeliZhitomirsky_PRL_2004_FrustratedBEC}.

Finally, we note that the above discussion yields a generalized
version of the ``zero scale universality'' behavior at $z=2$
critical point in one dimension~\cite{Sachdev_PRB_1994_ZeroScaleU}.
The difference is, unlike in the former case neither the numeric
values of the corresponding exponents (e.g., $m=1/2$, $\varphi=1$)
nor the functional form of $\mathcal{M}(x)$ are predefined.

\subsubsection{Quantum critical scaling: Experiment}

The manifestation of the experimentally accessible magnetization
quantum critical behavior is contained in Eq.~(\ref{EQ:MHscaling}).
To verify this relation and determine exponents $\varphi$ and $m$ we
studied the $H-T$ scaling of magnetization near saturation in the
axial geometry of \BaCdV. $M(H,T)$ data measured vs  applied field
at different temperatures are shown in Fig.~\ref{FIG:Mscale}(a).
Equation~(\ref{EQ:MHscaling}) suggests that all measurements are
expected to collapse onto a single curve if rescaled with
appropriate exponents. In order to determine the latter, for the
data in Fig.~\ref{FIG:Mscale}(a) we defined an empirical goodness of
overlap
criterion~\cite{PovarovSchmidiger_PRB_2015_DimpyScaling,*HaelgHuvonen_PRB_2015_NTENPscaling}.
Overall, this criterion is similar to a standard $\chi^2$ with the
only difference that the ``theoretical'' curve with respect to which
the deviation of datapoints is calculated is not predefined, but
empirically created on the fly at each iteration of the fit. A more
detailed description of the algorithm is contained in
Appendix~\ref{APP:scaling}.

Using $\mu_{0}H_{c}=3.95(2)$~T obtained in calorimetric measurements
as described below, we plot $\chi^2$ as a function of $m$ and
$1/\varphi$ in the inset in Fig.~\ref{FIG:Mscale}(b). The best
overlap is found for $m=0.8(1)$ and $1/\varphi=1.6(3)$, and results
in a spectacular data collapse shown in Fig.~\ref{FIG:Mscale}(b)
(main panel). The measured exponents are quite distinct from those
in the pure one-dimensional case, where $m=1/2$ and $\varphi=1$
~\cite{Sachdev_PRB_1994_ZeroScaleU,JeongRonnow_PRB_2015_CupzNscaling}.
Nonetheless, the observed exponent describing  the magnetization
cusp in the $T= 0$ limit [as given by Eq.~(\ref{EQ:MHcusp})] is the
same, namely, $m\varphi=0.5\pm0.15$, and agrees well with this
prediction made for the perfectly frustrated square
lattice~\cite{JackeliZhitomirsky_PRL_2004_FrustratedBEC}.

\begin{figure}
  \includegraphics[width=0.5\textwidth]{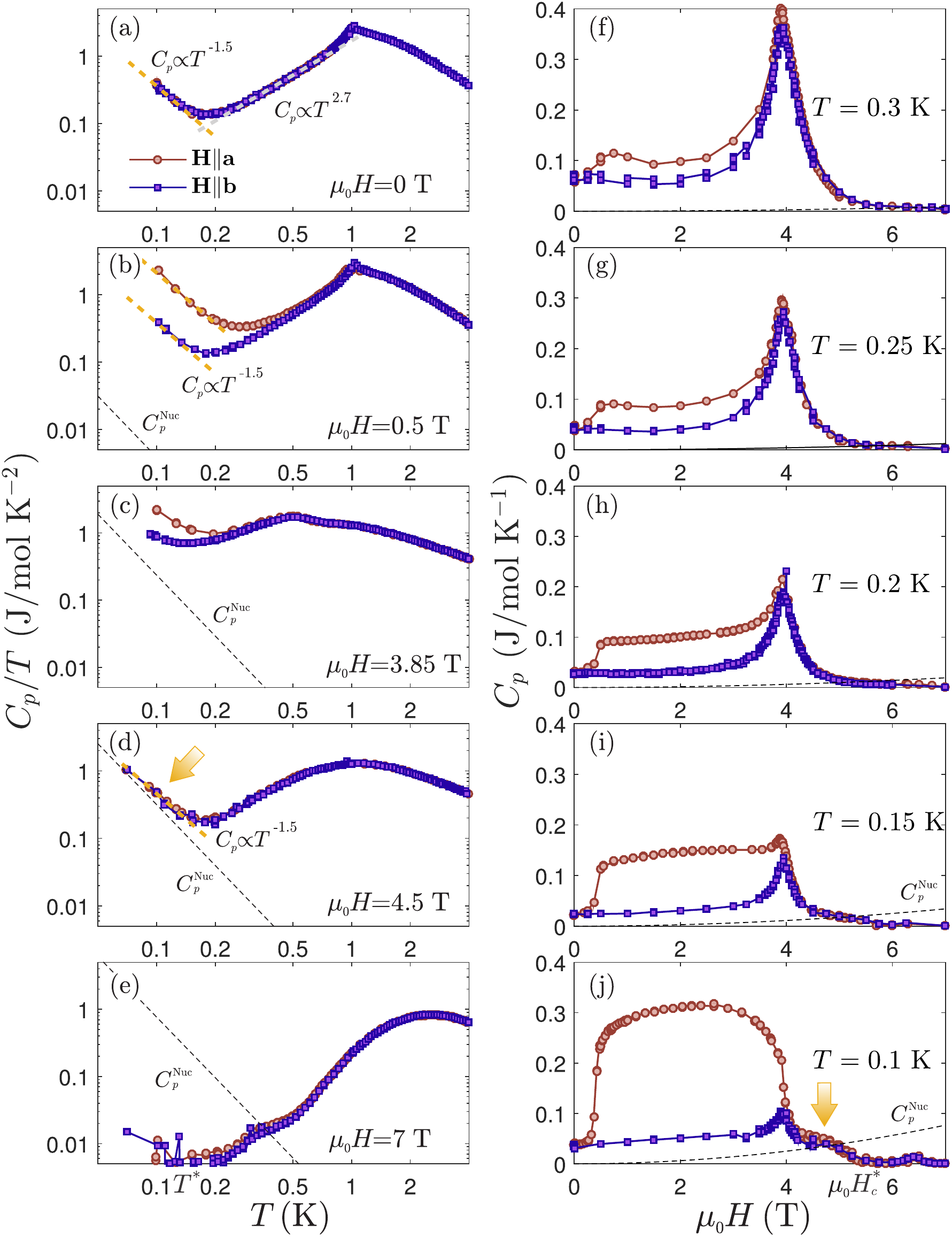}\\
  \caption{Low temperature specific heat in \BaCdV\ for axial and transverse geometries of the magnetic field.
  (a-e) $C_{p}(H,T)/T$ as the function of $T$ for different magnetic fields. Dashed lines
  show the power laws that can be identified in the data.
  (f-j) $C_{p}(H,T)$ at fixed temperature as the function of $H$. Arrows indicate excess specific
  heat appearing at low temperatures above the field-induced phase transition. In all the plots the dotted lines show the Zeeman
    effect based estimate of nuclear contribution,
Eqs.~(\ref{EQ:NucGen},\ref{EQ:NucContribs}).  Please note that the
panels on the left have logarithmic scale,
  while the panels on the right have linear scale. In
Appendix~\ref{APP:addendas} one can also see alternative ways of
plotting this data.}\label{FIG:Cp_panels}
\end{figure}

\subsection{Calorimetric studies}

\subsubsection{Specific heat measurements}

Further unusual behavior of \BaCdV\ was revealed by the specific
heat measurements. In both the
$\mathbf{H}\parallel\mathbf{a},\mathbf{b}$ orientations
zero-field-cooling data shows a pronounced lambda anomaly at
$T_{N}=1.05$~K followed by a power-law decrease in $C_{p}(T)/T$ as
shown in Fig.~\ref{FIG:Cp_panels}(a). This behavior is fully
consistent with the previously reported powder
data~\cite{NathTsirlin_PRB_2008_BaCdVO(PO4)2}.

Tracking the phase transition to reconstruct the $H-T$  phase
diagram is often easier in constant-$H$ scans, shown in
Figs.~\ref{FIG:Cp_panels}(f)-(j). However, these data reveal a
striking difference between axial and transverse geometries. The
first key result of our calorimetry studies is that in the axial
case, the field-induced transition becomes \emph{discontinuous} at
low temperatures. Above $T^{\ast}\simeq0.15$~K both geometries yield
a sharp $C_{p}(H)$ peak, marking a second-order transition
[Figs.~\ref{FIG:Cp_panels}(f)-\ref{FIG:Cp_panels}(h)] at a critical
field $H_c$. In the vicinity of $H_c$ and at all fields above it the
data for the two geometries are virtually indistinguishable. In
contrast, below $T^{\ast}$ the character of the anomaly in the axial
geometry changes. As shown in
Figs.~\ref{FIG:Cp_panels}(i),\ref{FIG:Cp_panels}(j) it rapidly
evolves from a peak to a steplike feature, similar to the step found
at the spin flop (a textbook example of discontinuous transition in
a magnet).

The second and perhaps the most important finding of our calorimetry
experiments is that there is an additional anomalous contribution to
specific heat at the lowest temperatures {\em above} $H_c$ in both
geometries. It can be seen in both constant-$T$
[Fig.~\ref{FIG:Cp_panels}(j)] and constant-$H$ scans
[Fig.~\ref{FIG:Cp_panels}(d)]. At 100~mK it persists as a plateau
all the way up to $\mu_{0}H_{c}^{\ast}\simeq5.2$~T, but vanishes at
higher fields [Fig.~\ref{FIG:Cp_panels}(e)].

\begin{figure}
  \includegraphics[width=0.5\textwidth]{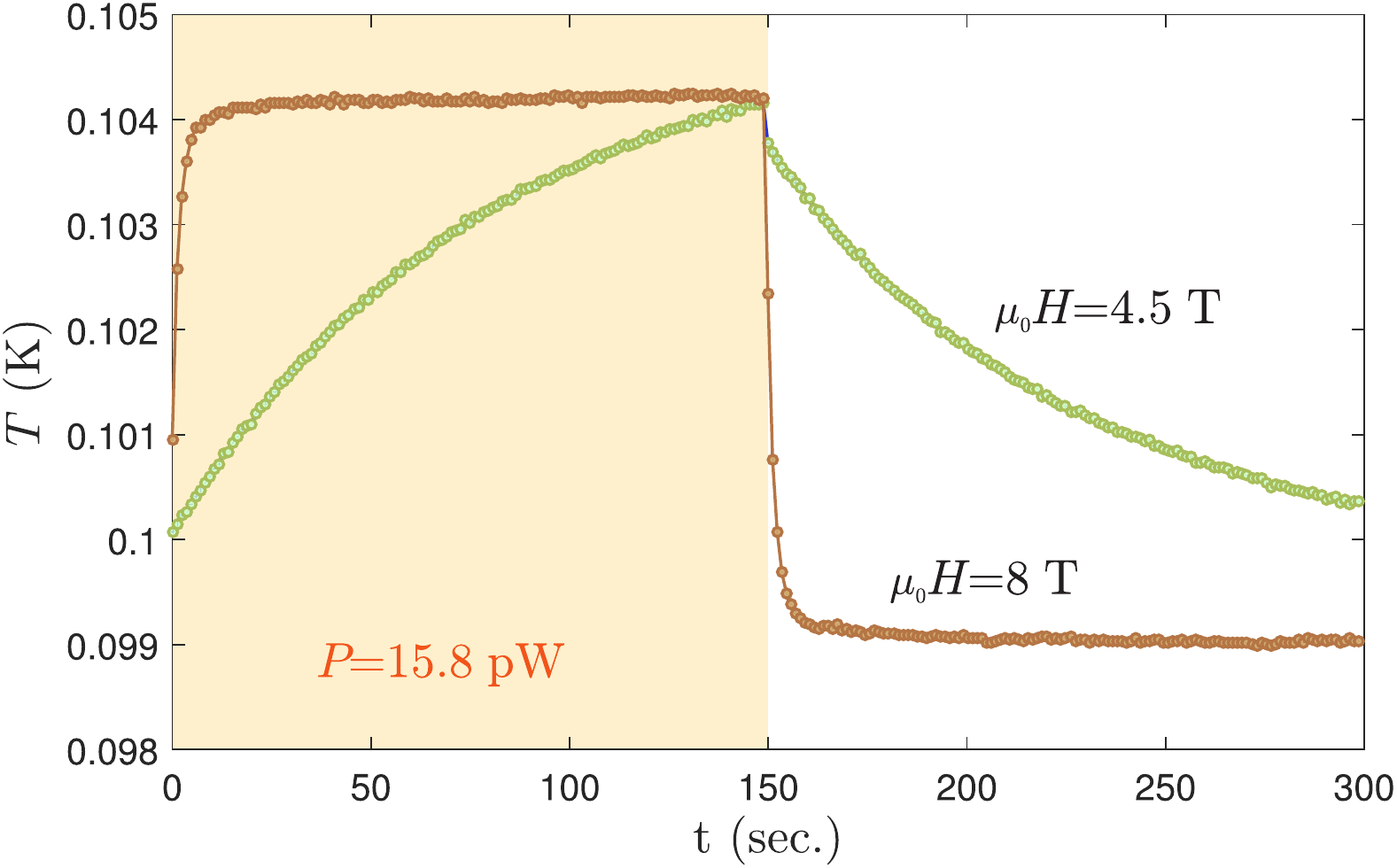}\\
  \caption{The raw specific heat data: relaxation curves taken at $\mu_{0}H=4.5$ and $8$~T in axial geometry.
  The two shown curves correspond to two single datapoints in Fig.~\ref{FIG:Cp_panels}(j).}\label{FIG:relaxations}
 \end{figure}

A very straightforward illustration of vanishing high-field specific
heat is shown in Fig.~\ref{FIG:relaxations} for the axial geometry
case. It demonstrates the relaxation curves obtained with the fixed
measurement time of $300$~s and magnitude of heating pulse
$P=15.8$~pW applied for $150$~s. The difference between the
measurements at fields of $4.5$~T$\gtrsim\mu_{0}H_{c}$ and
$8$~T$\gg\mu_{0}H_{c}$ at $T=100$~mK is apparent. At $4.5$~T the
relaxation curve indeed has the characteristic ``shark fin'' shape,
which means substantial specific heat present (as the rather long
measurement period is comparable to the characteristic relaxation
time). In contrast, at $8$~T after turning the heater on or off, in
a few seconds the system ends up in the stationary regime. This is
the clear signature of very short relaxation time and almost absent
specific heat in the sample.

\subsubsection{Magnetocaloric effect measurements}

\begin{figure}
  \includegraphics[width=0.5\textwidth]{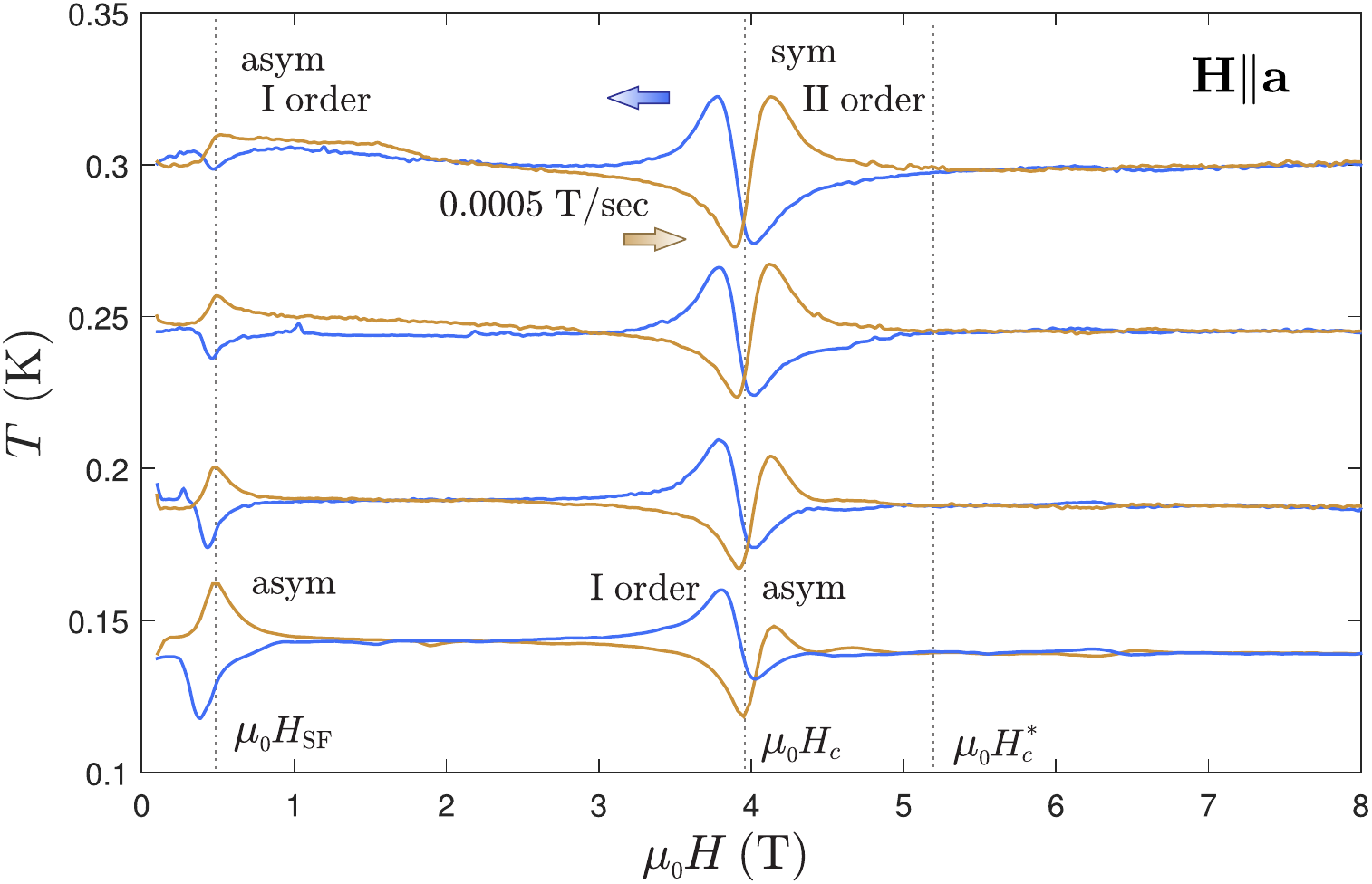}\\
  \caption{Magnetocaloric effect in \BaCdV\ at low temperatures in the axial geometry. The $T(H)$ dependencies taken at different temperatures
  with the field sweeping rate of $\pm5\times10^{-4}$~T/s.}\label{FIG:MCE}
\end{figure}

The discontinuous character of the low-temperature field-induced
transition in the axial case is also confirmed by the measurements
of the magnetocaloric effect. Utilizing the same experimental setup
as for the relaxation calorimetry, we monitor the sample temperature
during slow magnetic field sweeps, while keeping the heat bath
temperature constant. In this so-called equilibrium
regime~\cite{AczelKohama_PRL_2009_Sr3Cr2O8BEC} the excess thermal
power created due to the sample's entropy change is balanced by the
temperature gradient between the sample and the bath across the weak
heat link. The evolution of the resulting sample's $T(H)$ curves for
up and down magnetic field sweeps is shown in Fig.~\ref{FIG:MCE}.
The first-order spin-flop transition manifests itself as a highly
asymmetric peaklike feature at all the temperatures. This is a
direct consequence of the entropy discontinuity. In contrast, at
elevated temperatures the magnetocaloric anomaly at $H_{c}$  is very
symmetric, as it should be for a continuous
transition~\cite{GarstRosch_PRB_2005_MagnetocaloricTheory,SchmidtThalmeier_PRB_2007_J1J2squareMCeffect}.
However, below around $T^{\ast}$ this anomaly rapidly becomes rather
asymmetric as well, confirming the change of the transition type.

\subsubsection{Possible nuclear specific heat contributions}

Although observations of divergent low-temperature specific heat due
to nuclear magnetism with extremely low-energy scale are common,
below we will show that the simple picture is qualitatively
inconsistent with the present data.

The simplest model of nuclear specific heat assumes the energy
levels in the spinful nuclei being split due to Zeeman effect. Then
the nuclear contribution is approximately given as:

\begin{equation}\label{EQ:NucGen}
    C_{p}^{\text{Nuc}}(T,H)=\mathfrak{A}\left(\frac{\mu_{0}H}{T}\right)^2.
\end{equation}

The material-dependent amplitude coefficient is calculated as
follows:

\begin{equation}\label{EQ:NucContribs}
    \mathfrak{A}=\sum\limits_{i}\mathfrak{A}_{i}=\sum\limits_{i}N_{A}n_{i}\alpha_{i}\frac{I_{i}(I_{i}+1)(\gamma_{i}\hbar)^{2}}{3k_{B}}.
\end{equation}

The summation goes through all the spinful types of nuclei present
in the material; $n_i$ is the stoichiometric coefficient in the
chemical formula and $\alpha_i$ is the abundance of the particular
isotope. The data on the isotopes abundance, nuclear spins $I_i$,
and corresponding nuclear gyromagnetic ratios $\gamma_i$ are found
in Ref.~\cite{Mason_1987_MultNMR}, for example. The isotope data and
the corresponding contribution to the nuclear specific heat
prefactor relevant to \BaCdV\ are summarized in
Table~\ref{TAB:Isotopes}. The overall $C_{p}^{\text{Nuc}}(T,H)$
prefactor  is estimated as
$\mathfrak{A}=1.5671\times10^{-5}$~J~K/mol~T$^{-2}$, with about 80\%
of it stemming from the magnetic ion $^{51}$V having nuclear spin
$I=7/2$. This means that for the consistent description of
$C_{p}^{\text{Nuc}}$ the quadrupolar splitting and hyperfine
interactions on the $^{51}$V site also need to be taken into
account. These parameters are unknown at the moment, and therefore
Eqs.~(\ref{EQ:NucGen},\ref{EQ:NucContribs}) should be seen only as
the crude estimate of possible effect magnitude. As one can see from
Fig.~\ref{FIG:Cp_panels}, the low-temperature specific heat in
\BaCdV\ is completely at odds with this simple estimation.

 Nonetheless,
since nuclear spin $I=7/2$ is also carried by the magnetic $S=1/2$
$^{51}$V$^{4+}$ ions, some complex behavior induced by hyperfine
coupling close to the quantum critical point can not be fully ruled
out. There are some
experimental~\cite{Ronnow_Science_2005_nuclearQCP} and
theoretical~\cite{TsvelikZaliznyak_PRB_2016_NuclearNecklace} studies
of hyperfine coupled settings, but not for the strongly frustrated
2D case.

\begin{table}[t]
\begin{tabular}{ c r c r c r c r c r c }
  \hline\hline
    Isotope & &$\alpha_i$ & &$n_i$ & & $\gamma_i$ (rad/s T$^{-1}$) & & $I_i$ & & $\mathfrak{A}_i$ (J K/mol T$^{-2}$) \\
    \hline
  $^{135}$Ba & &0.06590& & 1& & $2.6755\times10^{7}$ & & 3/2 & & $2.8604\times10^{-8}$ \\
  $^{137}$Ba & &0.11320& & 1& & $2.9930\times10^{7}$ & & 3/2 & & $6.1488\times10^{-8}$ \\
  $^{111}$Cd & &0.12750& & 1& & $-5.7046\times10^{7}$ & & 1/2 & & $5.0318\times10^{-8}$ \\
  $^{113}$Cd & &0.12260& & 1& & $-5.9609\times10^{7}$ & & 1/2 & & $5.2829\times10^{-8}$ \\
  $^{50}$V &   &0.00240& & 1& & $2.6721\times10^{7}$ & & 6 & & $1.1638\times10^{-8}$ \\
  $^{51}$V &   &0.99760& & 1& & $7.0492\times10^{7}$ & & 7/2 & & $1.2625\times10^{-5}$ \\
  $^{31}$P &   &1.00000& & 2& & $1.0839\times10^{8}$ & & 1/2 & & $2.8497\times10^{-6}$ \\
  $^{17}$O &   &0.00037& & 9& & $-3.6280\times10^{7}$ & & 5/2 & & $6.2013\times10^{-9}$ \\
  \hline\hline
\end{tabular}
\caption{The relevant isotope data (from
Ref.~\cite{Mason_1987_MultNMR}) and corresponding calculated
contribution [Eq.~(\ref{EQ:NucContribs})] to the nuclear specific
heat due to Zeeman splitting.}\label{TAB:Isotopes}
\end{table}

\subsubsection{Magnetic phase diagram of \BaCdV}

Leaving aside the certainly exotic scenario of interplay between the
nuclear and electronic spins, we face the conclusion that the found
excess specific heat is of purely electron spin origin. Apart from
either electronic or nuclear spins no other degrees of freedom may
give a field-dependent contribution to the specific heat of an
insulating material at these low temperatures. We conclude that in
\BaCdV at the lowest temperatures $H_c$ does {\em not} correspond to
the full saturation. Indeed, the latter would open a Zeeman gap in
the spectrum and suppress any magnetic specific heat. Instead, $H_c$
indicates the appearance of a {\em new quantum regime with
substantial low-energy fluctuations}.

\begin{figure}
  \includegraphics[width=0.5\textwidth]{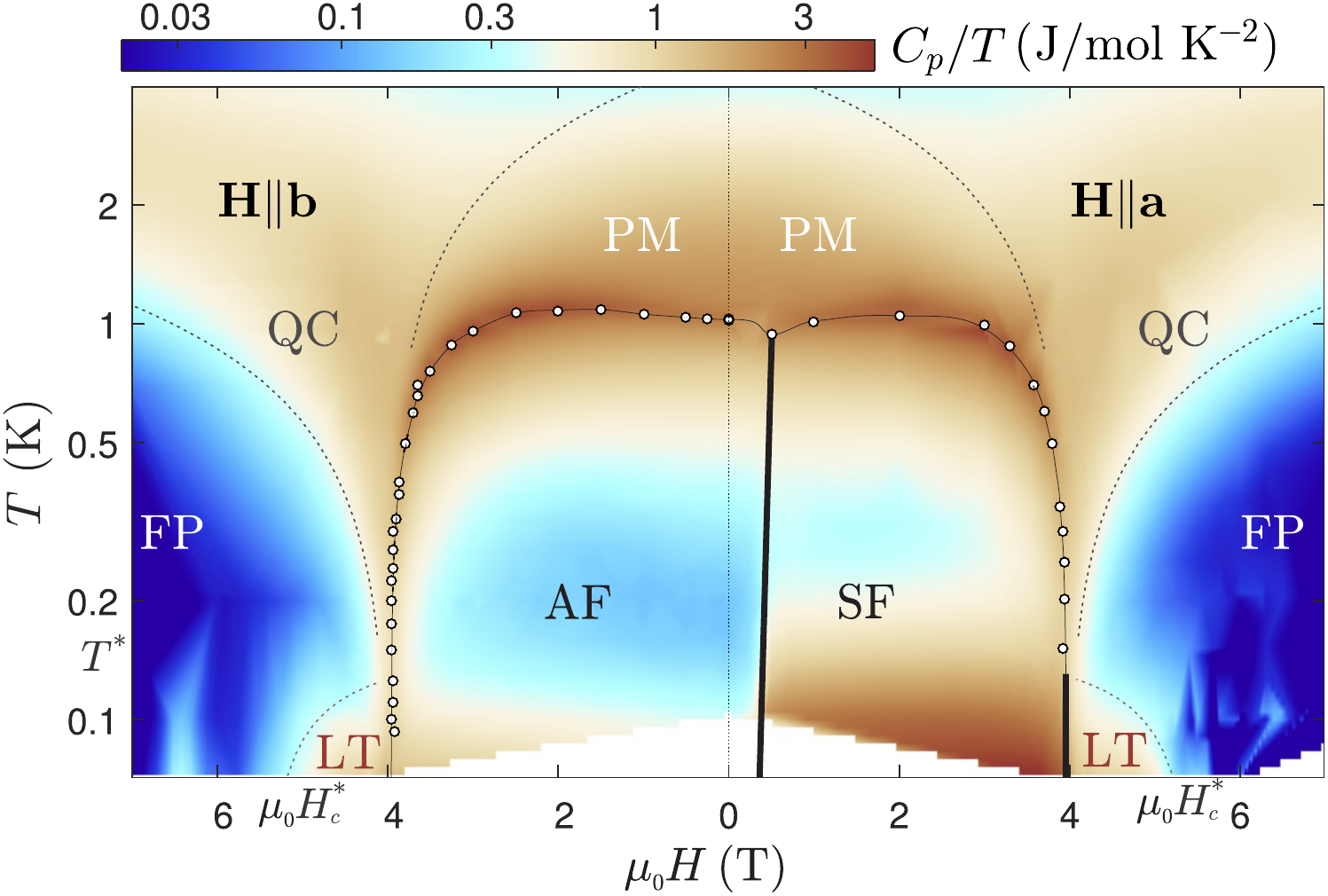}\\
  \caption{Magnetic phase diagram for $\mathbf{H}\parallel\mathbf{a,b}$. The background shows the false color map
  of $C_{p}(T,H)/T$, thin and thick black solid lines represent the phase transitions (of second or first order
  correspondingly),
  and gray dashed lines mark crossovers.
  Points are the ordered phase boundary data obtained from $C_{p}$ anomalies. The phases are as follows: PM,
  paramagnetic;
  FP, fully polarized;
  AF, antiferromagnetic; SF, antiferromagnetic after the spin flop;
  QC, quantum critical regime; LT, unconventional low temperature regime. Crossover lines marking
  the QC regime follow $T\propto|H-H_{c}|^{\varphi}$ with the same crossover exponent $\varphi$ found from scaling Eq.~(\ref{EQ:MHscaling}).
  }\label{FIG:AB_Cpmap}
\end{figure}

The magnetic $H-T$ phase diagram of \BaCdV in
Fig.~\ref{FIG:AB_Cpmap} summarizes the findings. We distinguish
conventional paramagnetic (PM), field polarized (FP), and AF states
(and its post-spin-flop version SF). At intermediate temperatures a
quantum critical (QC) regime  is observed above $H_{c}$. The new
low-temperature field-induced states are labeled as LT. While they
are separated from the ordered states by obvious phase transitions,
their finite-$T$ boundaries cannot be clearly identified in our
calorimetry data. Thus, we simply identify the crossover line below
which the anomalous behavior becomes pronounced. The appearance of
the LT regime, already intriguing on its own, becomes especially
interesting if considered in the context of predictions made for the
spin-nematics.

\subsubsection{Observed anomalies in context of spin-nematics}

One can find interesting possible connections of the observed
anomalies to the expected behavior of the two-dimensional $S=1/2$
spin-nematic materials. First, we would like to note that the
location of the LT regimes is \emph{in principle} consistent with
the expectations for the spin-nematic state in the frustrated square
lattice model. The anomalous specific heat found in \BaCdV\ samples
is endemic to the very low temperatures compared to the typical
interactions of the order of a few kelvin in the material.
Nonetheless, this is indeed the temperature range in which the
anomalous behavior due to magnons pairing up is expected to take
place from the theory point of view. Exact diagonalization studies
of the frustrated model with $J_{2}/J_{1}=-0.4$ suggest significant
nematic-type contributions to the specific heat to occur at
temperatures order of magnitude lower than typical $J$'s in the
system~\cite{ShannonMomoi_PRL_2006_J1J2squarecircle}. In the case of
\BaCdV\ the relevant energy scale can be suppressed even further
below $0.1J_{1}\simeq0.3$~K, as the system deviates from the
idealized $J_{2}/J_{1}=-0.4$ zero-field case. This looks very
consistent with our present observations shown in
Fig.~\ref{FIG:Cp_panels}(d). At the same time the spin-nematic
precursor behavior may in principle be present at any field below
the true saturation point
--- as long as there are fluctuating transverse spin components. There are indications that spin fluctuations associated with the new
low-temperature high-field phases are present already in the ordered
states. There too we find anomalous contributions to specific heat
below the crossover temperature $T^{\ast}\simeq0.15$~K
[Figs.~\ref{FIG:Cp_panels}(a)-\ref{FIG:Cp_panels}(c)]. They roughly
follow $C_{p}(T)\propto T^{-1.5}$ and are particularly strong in the
spin-flop state. This suggests their transverse character. The field
$H_{c}^{\ast}$ at which the spin fluctuations vanish is consistent
with the effective magnetic energy scale of the material
$J_{\text{eff}}=\sqrt{J_{1}^{2}+J_{2}^2}$~\cite{Shannon_EPJB_2004_GenericJ1J2theory}.

A second interesting observation is related to the boundary between
conventional antiferromagnetic and LT phases. As shown above, in the
axial geometry the new high-field state is entered from the
spin-flop AF phase through a {\em discontinuous transition}.
Incidentally, this is exactly the type of behavior expected for the
spin-nematic phase predicted to emerge just below full
saturation~\cite{SmeraldShannon_PRB_2016_nematicNMR}. The AF and
spin-nematic phases have competing order parameters, and therefore
the transition between them has to be first order. In the transverse
geometry, the spin-nematic phase should not exist in a field due to
a lack of axial symmetry~\cite{Zhang_PRB_2017_NematicAnisotropy}.
However, this is not supposed to impede the associated fluctuations
completely. While strong spin fluctuations persist irrespective of
field orientation in \BaCdV, they may result in nematic order only
in the axial geometry. This may explain why the transition at $H_c$
remains continuous in the transverse case and becomes discontinuous
in the axial
--- the only case expected to support the nematic long-range
ordering.

To summarize, it is very tempting to consider the observed anomalous
regime LT as the precursor of the true spin-nematic long-range
order. This would simultaneously explain the small energy scale
associated with the new state as well as the
field-direction-dependent transition type. However, at this stage we
still cannot fully rule out the possibility of interference between
the electronic and the nuclear spins going beyond the simple model
described by Eq.~(\ref{EQ:NucContribs}).

\section{Conclusions}

The high hopes for finding the unconventional magnetism in the
frustrated $S=1/2$ square lattice magnet \BaCdV\ appear to be well
justified. In addition to the experimentally quantified
``dimensionality reduction'' effect serving as the indicator of
strong frustration we have also found anomalously strong
contributions to the specific heat in the vicinity of saturation
field at lowest temperatures. Although the possibility of their
origin from the interplay of electronic and nuclear magnetism is not
yet fully ruled out, these anomalies show qualitative consistency in
the order of phase transition and energy scale with the predicted
spin-nematic behavior. Future efforts aimed at understanding the
origins of the novel regime will have to specifically focus on the
lowest possible temperatures.

\acknowledgments

This work was supported by Swiss National Science Foundation,
Division II. We would like to thank Prof. Oleg Starykh (University
of Utah) for enlightening discussions, Stanislaw Galeski and Dominic
Blosser (ETH Z\"{u}rich) for help with the low-temperature
magnetocaloric measurements, and Dr. Severian Gvasaliya (ETH
Z\"{u}rich) for technical assistance.

\appendix

\section{Definition of scaling $\chi^{2}$ criterion} \label{APP:scaling}
\begin{figure}
  \includegraphics[width=0.5\textwidth]{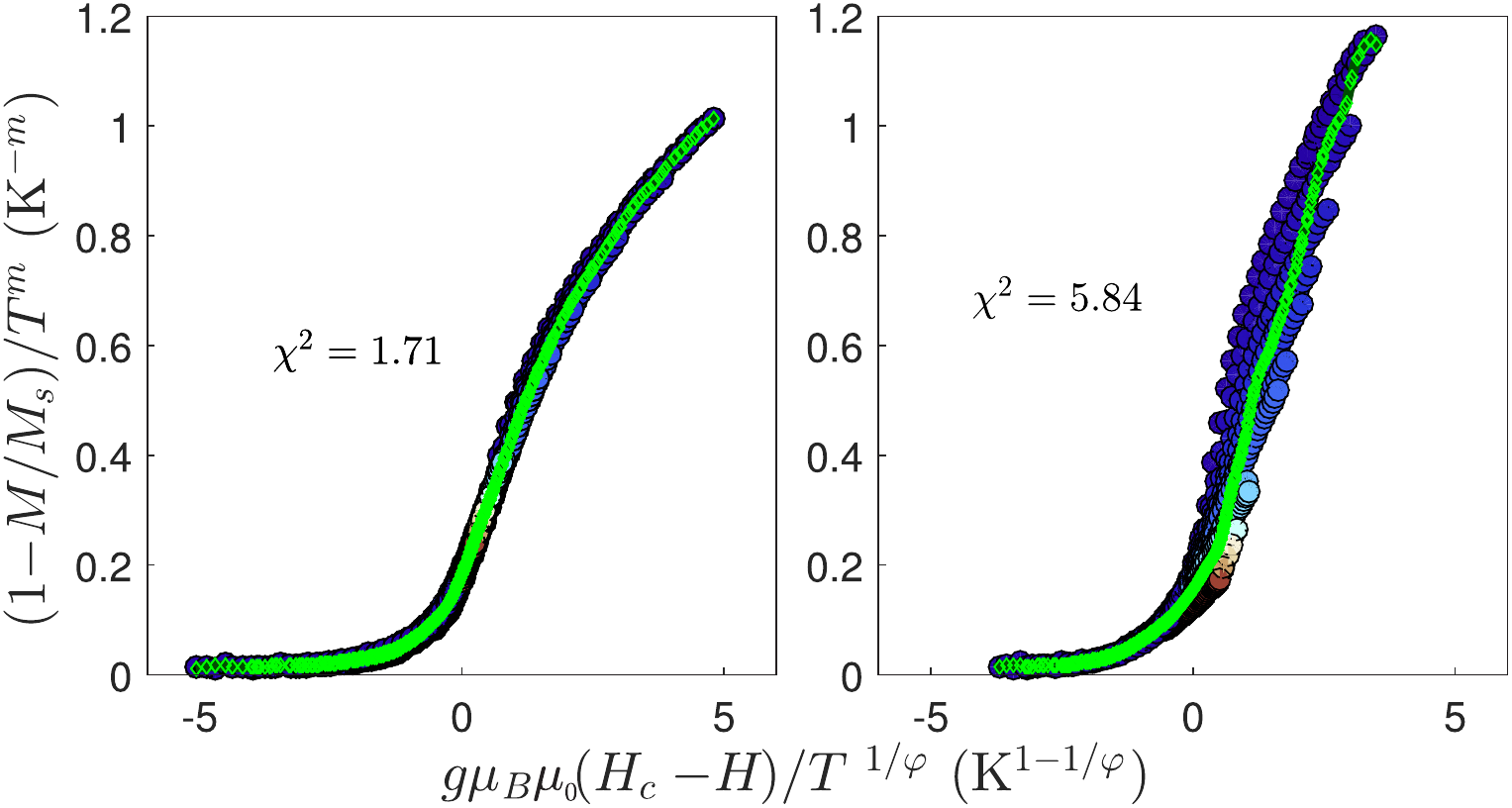}\\
  \caption{The scaling analysis of the magnetization data (see Fig.~\ref{FIG:Mscale} of the main text). Left: optimal scaling exponents $1/\varphi=1.55$, $m=0.76$.
  Right: non-optimal scaling exponents $1/\varphi=1$, $m=1$.
  Green curve represents the empirical data interpolation with respect to which the $\chi^2$
  costs are calculated.}\label{FIG:X2}
\end{figure}

The hypothesis that is being tested for the magnetization data
present in Fig.~\ref{FIG:Mscale} of the main text is that it follows
the universal behavior in the vicinity of $H_{c}$:

\begin{equation}
\label{EQ:MHscalingSM}
    1-M(T,H)/M_{\text{sat}}=T^{m}\mathcal{M}\left(\frac{g\mu_{0}\mu_{B}(H_{c}-H)}{T^{1/\varphi}}\right).
\end{equation}

This means that for correctly chosen exponents $m$ and $\varphi$ the
set of datapoints $X=\frac{g\mu_{0}\mu_{B}(H-H_{c})}{T^{1/\varphi}}$
and $Y=[1-M(H)/M_{\text{sat}}]/T^{m}$ should lie close to some
hypothetical curve. In principle, if this hypothetical curve
$Y_{0}(X)$ is known, the problem of calculating the abstract
``goodness of overlap'' can be reduced to the very standard problem
of calculating  of the ``goodness of fit'' of the data $Y(X)$ by
theory  $Y_{0}(X)$.

The key idea in the present approach, where no \emph{a priori}
scaling curve is postulated, is to construct $Y_{0}(X)$ ``on the
fly'' based on the current $Y(X)$ data. This is achieved by
interpolating the scattered $Y(X)$ with cubic splines. It guarantees
the smoothness of the resulting curve and at the same time gives a
bit more flexibility than polynomial interpolation used, e.g., in
Ref.~\cite{JeongRonnow_PRB_2015_CupzNscaling} in a similar
situation. The examples of such an empirical interpolation curve for
cases with good and poor choices of scaling exponents are shown in
Fig.~\ref{FIG:X2}.

A remark needs to be made regarding the normalization of cost
function in the case described above. The $\chi^{2}$ value is
usually normalized with the number of degrees of freedom, which is
typically the number of datapoints. However, in the present
situation individual degrees of freedom are rather represented by
the individual $M(H)$ scans at fixed temperatures. For any
separately taken scan the interpolation procedure would by
definition provide an ideal overlap with the ``empirical curve'',
and it is the optimization in the presence of multiple such datasets
that constitutes the essence of the procedure. Then the cost
function, being the equivalent of a standard normalized
error-bar-weighted $\chi^{2}$ is calculated as follows:

\begin{equation}\label{EQ:X2}
    \chi^{2}=\frac{1}{N_{\text{Datasets}}-1}\sqrt{\sum\limits_{X_{i}}\left(\frac{Y(X_{i})-Y_{0}(X_{i})}{\Delta Y(X_{i})}\right)^{2}}.
\end{equation}

\section{Addenda in the specific heat measurements}
\label{APP:addendas}

\begin{figure*}
  \includegraphics[width=0.8\textwidth]{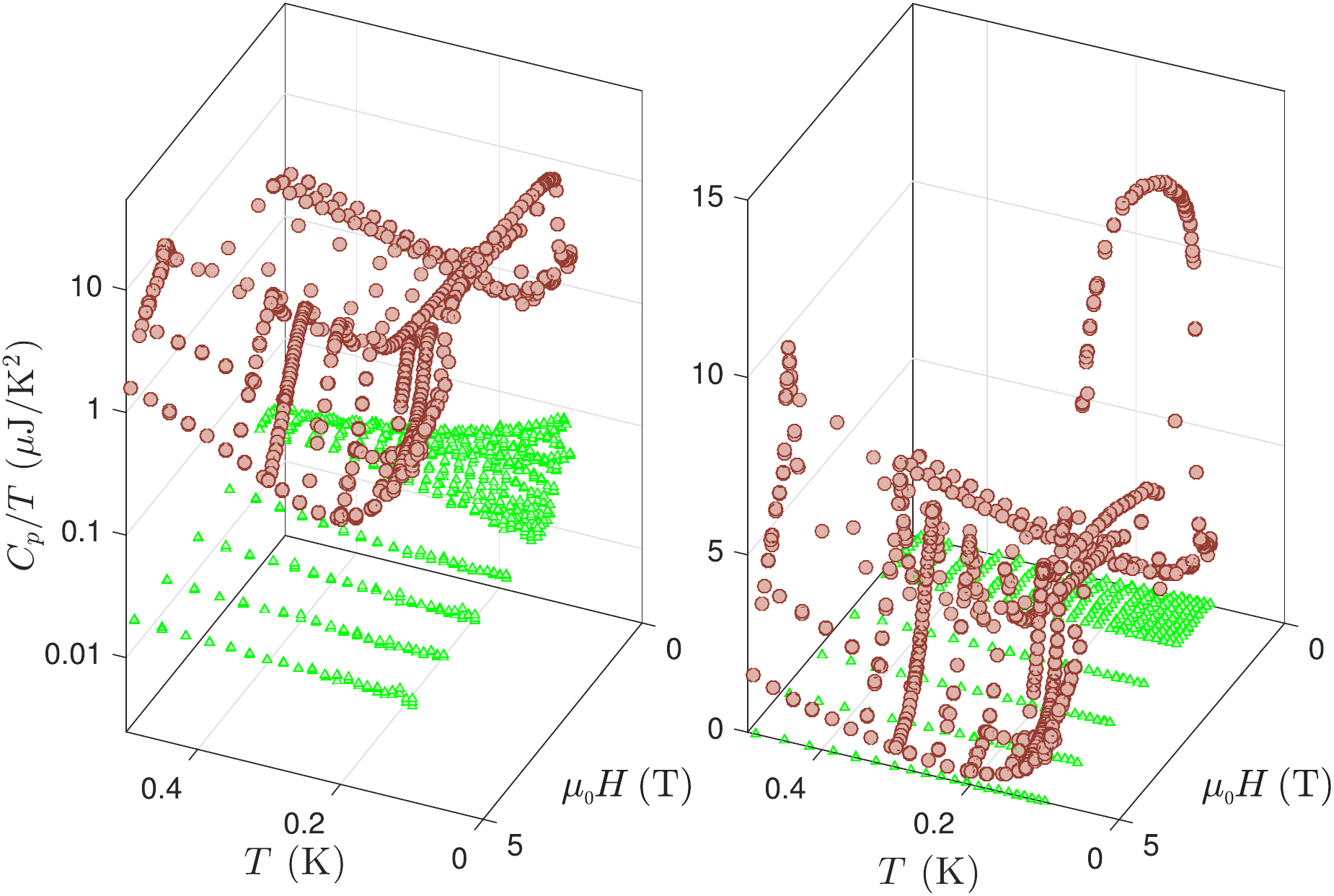}\\
  \caption{Low temperature part of addenda specific heat (triangles) and total specific heat (circles) with \BaCdV\ sample mounted in axial geometry. Left panel shows $C_{p}/T$ data on semilogarithmic scale, right panel shows the data on linear scale.}\label{FIG:Maddenda}
\end{figure*}

Finally, we would like to present a proof that the observed
anomalous specific heat is sample related. One simple consideration
is that the addenda contribution is somewhat different in both
$\mathbf{H}\parallel\mathbf{a}$ and $\mathbf{H}\parallel\mathbf{b}$
cases (as different amounts of grease and a different piece of
silver foil holder was used), while the observed extra specific heat
is well matched. But even more valuable is the direct comparison,
given in Fig.~\ref{FIG:Maddenda} for the
$\mathbf{H}\parallel\mathbf{a}$ setup. One can see that the
background specific heat contribution is very small. Apart from a
tiny Shottky anomaly close to $H=0$ it is dominated by $C_{p}\propto
T$ linear specific heat of the silver foil.

\bibliography{d:/The_Library}

\end{document}